\documentclass[epj]{svjour}

\usepackage{graphicx}
\usepackage{fancyhdr}
\def\makeheadbox{{
 }}
\def\mytitle{My title} 
\def\myauthors{My name}  
\def\mytype{My type of session}
\def\mysession{My session}

\def\mytitle{CP violation in SUSY} 
\def\myauthors{Sabine Kraml}    
\def\mytype{Review}
\def\mysession{\myauthors}

\begin{document}
\title{CP violation in SUSY}
\author{Sabine Kraml\inst{1,2}
\thanks{\emph{Email:} sabine.kraml@cern.ch}%
}                     
\institute{CERN PH-TH, 1211 Geneva 23, Switzerland \and LPSC, 53 Av des Martyrs, 38026 Grenoble, France}
%
\date{}
\abstract{CP violation in supersymmetric models is reviewd with focus on explicit CP 
violation in the MSSM. The topics covered in particular are CP-mixing in the Higgs 
sector and its measurement at the LHC, CP-odd observables in the gaugino sector at the ILC, 
EDM constraints, and the neutralino relic density.     
} 
\maketitle
%

\section{Introduction} \label{intro}

Test of the discrete symmetries, charge conjugation C, parity P, and time-reversal T,
have played an important role in establishing the structure of Standard Model (SM).  
In particular, CP violation has been observed in the electroweak sector of the SM in 
the $K$ and $B$ systems. It is linked to a single phase in the unitary Cabbibo-Kobayashi-Maskawa 
(CKM) matrix describing transitions between the three generations of quarks; 
see e.g.\ \cite{Buras:2005xt} for a detailed review. It is important to note that this 
source of CP violation is strictly flavour non-diagonal. 

The strong sector of the SM also allows for CP violation through a dimension-four 
term $\theta G\tilde G$, which is of topological origin. Such a term would lead to 
flavour-diagonal CP violation and hence to electric dipole moments (EDMs).
The current experimental limits on the EDMs of atoms and neutrons~\cite{Tl,Hg,n}
\begin{equation}
\begin{array}{rl}
  |d_{\rm Tl}| < 9 \times 10^{-25}\, e\, {\rm cm}     &\quad \rm (90\%\, C.L.) \\ 
  |d_{\rm Hg}| < 2   \times 10^{-28}\, e\, {\rm cm} &\quad \rm (95\%\, C.L.) \\ 
  |d_n|  <  6\times 10^{-26}\, e\, {\rm cm}             &\quad \rm (90\%\, C.L.) \\ 
\end{array}\label{eq:edmconstr}
\end{equation}
however constrain the strong CP phase to $|\theta|<10^{-9}$! 
A comprehensive discussion of this issue can be found in~\cite{Pospelov:2005pr}.
While $\theta$ appears to be extremly tuned, the CKM contribution to the EDMs is several 
orders of magnitude below the experimntal bounds, e.g.\ $d_n^{\rm CKM}\sim 10^{-32}\, e\, {\rm cm}$.
Therefore, while providing important constraints, the current EDM bonds still leave ample room for 
new sources of CP violation beyond the SM. 

Such new sources of CP violation are indeed very interesting in 
point of view of the observed baryon asymmetry of the Universe 
\begin{equation}
   \eta = \frac{n_B-n_{\bar B}}{n_\gamma} = (6.14\pm 0.25)\times 10^{-10}
\end{equation}
with $n_B$, $n_{\bar B}$ and $n_\gamma$ the number densities of 
baryons, antibaryons and photons, respectively; see \cite{Dine:2003ax,Cline:2006ts}
for recent reviews. 
The necessary ingredients for baryogenesis \cite{Sakharov:1967dj}
i) baryon number violation, ii) C and CP violation and iii) departure from equilibrium
are in principle present in the SM, however not with sufficient strength. In particular, 
the amount of CP violation is not enough. This provides a strong motivation 
to consider CP violation in extensions of the SM, as reviewed e.g.\ in \cite{Ibrahim:2007fb}.

In general, CP violation in extensions of the SM can be either explicit or spontaneous. 
Explicit CP violation occurs through phases in the Lagrangian, which cannot be rotated 
away by field redefinitions. This is the standard case in the MSSM, on which I will concentrate 
in the following.
Spontaneous CP violation, on the other hand, occurs if an extra Higgs field develops a 
complex vacuum expectation value. This can lead to a vanishing $\theta$ term as well 
as to a complex CKM matrix.  Spontaneous CP violation is a very interesting and elegant idea, 
but difficult to realize in SUSY and obviously not possible in the MSSM (where the Higgs potential 
conserves CP). There has, however, been very interesting 
new work on left-right symmetric models and SUSY GUTs. For instance, models based on 
supersymmetric SO(10) may provide a link with the neutrino seesaw and leptogenesis.
I do not follow this further in this talk but refer to \cite{Ibrahim:2007fb} for a review. 

\section{CP violation in the MSSM}\label{sec:mssm} 

In the general MSSM, the gaugino mass parameters $M_i$ ($i=1,2,3$), the higgsino mass parameter $\mu$, 
and the trilinear couplings $A_f$ can be complex,
\begin{equation}
  M_i = |M_i|\,e^{i\phi_i}, \quad 
  \mu = |\mu|\,e^{i\phi_\mu}, \quad 
  A_f = |A_f|\,e^{i\phi_f},
\label{eq:phases}
\end{equation}
(assuming $B\mu$ to be real by convention) thus inducing explicit CP violation in the model. 
Not all of the phases in eq.~(\ref{eq:phases}) are, however, physical. The physical combinations 
indeed are ${\rm Arg}(M_i\mu)$ and ${\rm Arg}(A_f\mu)$. They can 
\begin{itemize}
\item affect sparticle masses and couplings through their mixing,
\item induce CP mixing in the Higgs sector through radiative corrections,
\item influence CP-even observables like cross sections and branching ratios, 
\item lead to interesting CP-odd asymmetries at colliders.
\end{itemize}
Non-trivial phases, although constrained by EDMs, can hence significantly 
influence the collider phenomenology of Higgs and SUSY particles, and as 
we will see also the properties of neutralino dark matter. 

Let me note here that CP violation in the MSSM alone is a large field with 
a vast amount of literature; it is essentially impossible to give a complete 
review in 25\,min.  I will hence not try a {\it tour de force} but rather present 
some selected examples, and I apologize to those whose work is not 
mentioned here. This said, let us begin with the MSSM Higgs sector:
 
\subsection{Higgs-sector CP mixing}\label{sec:higgs}

The neutral Higgs sector of the MSSM consists  in principle of two CP-even states, 
$h^0$ and $H^0$, and one CP-odd state, $A^0$. 
Complex parameters, eq.~(\ref{eq:phases}), here have a dramatic effect, inducing a mixing between
the three neutral states through loop corrections~\cite{Pilaftsis:1998dd,Demir:1999hj,Pilaftsis:1999qt}. 
The resulting mass eigenstates $H_1,\,H_2,\,H_3$ (with $m_{H_1}<m_{H_2}<m_{H_3}$ by convention) are 
no longer eigenstates of CP. Owing to the large top Yukawa coupling, 
the largest effect comes from stop loops, 
with the size of the CP mixing proportional to \cite{Choi:2000wz}
\begin{equation}
  \frac{3}{16\pi^2}\,\frac{\Im m(A_t\mu)}{m_{\tilde t_1}^2-m_{\tilde t_2}^2}\,.
\end{equation}

\begin{figure}
\includegraphics[width=0.48\textwidth,angle=0]{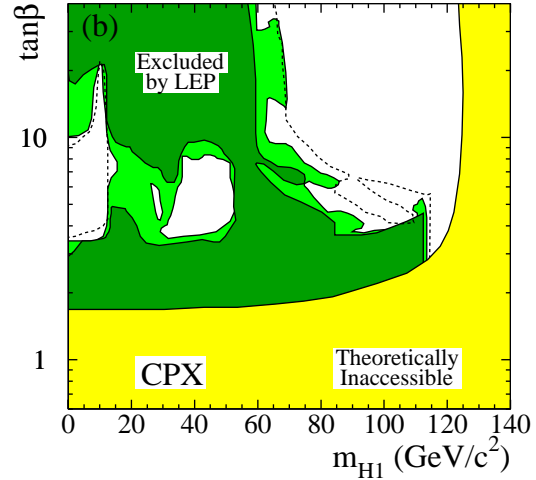}
\caption{LEP limits in the CPX scenario, from \cite{Abbiendi:2006cr}.}
\label{fig:leplimits}       
\end{figure}

\begin{figure}
\hspace{6mm}\includegraphics[width=0.44\textwidth,angle=0]{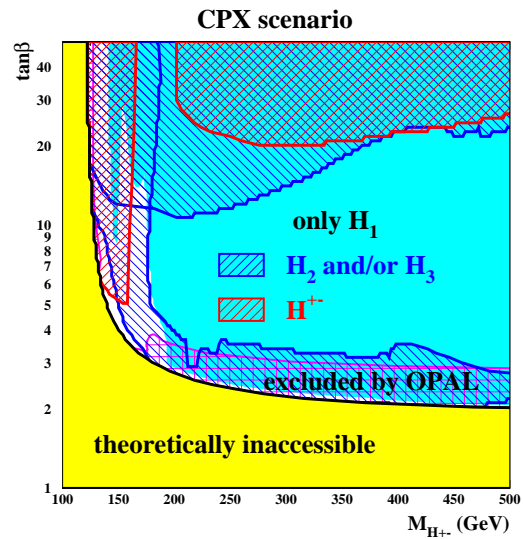}
\caption{ATLAS discovery potential for Higgs bosons in the CPX scenario, from \cite{Schumacher:2004da}.}
\label{fig:atlaspot}       
\end{figure}

CP mixing in the Higgs sector can change the collider phenomenology 
quite substantially. For example, it is possible for the lightest Higgs boson to
develop a significant CP-odd component such that its coupling to a pair of 
vector bosons becomes vanishingly small. 
This also considerably weakens the LEP bound on the lightest Higgs boson 
mass~\cite{Abbiendi:2006cr}, as illustrated in Fig.~\ref{fig:leplimits},  
which shows the LEP exclusions at 95\% CL (medium-grey or light-green) and 99.7\% CL 
(dark-grey or dark-green) for the CPX scenario with maximal phases; the top mass 
is taken to be $m_t=174.3$~GeV.
The CPX scenario~\cite{Carena:2000ks} is the default benchmark scenario for 
studying CP-violating Higgs-mixing phenomena. It is defined as  
\begin{eqnarray}
&& M_{\tilde{Q}_3} = M_{\tilde{U}_3} = M_{\tilde{D}_3} =
     M_{\tilde{L}_3} = M_{\tilde{E}_3} = M_{\rm SUSY}\,, \\
&& |\mu|=4\,M_{\rm SUSY}\,,~ |A_{t,b,\tau}|=2\,M_{\rm SUSY} \,,~  |M_3|=1~{\rm TeV}\,.
     \nonumber
\label{eq:CPXdef}
\end{eqnarray}
The free parameters are $\tan\beta$, the charged Higgs-boson pole mass
$M_{H^\pm}$, the common SUSY scale $M_{\rm SUSY}$, and the CP phases. 
Typically one chooses $\phi_\mu=0$, which leaves $\phi_{t,b}$ and $\phi_3$ 
as the relevant ones. 
The ATLAS discovery potential \cite{Schumacher:2004da} for Higgs bosons in the CPX scenario  
with $\phi_{t,b,3}=\pi/2$ is shown in Fig.~\ref{fig:atlaspot}. As can be seen, also  
here there remains an uncovered region at small $\tan\beta$ and small Higgs masses, 
comparable to the holes at small $m_{H_1}$ in Fig.~\ref{fig:leplimits}, 

An overview of the implications for Higgs searches 
at different colliders is given in~\cite{Godbole:2004xe}, 
and a review of MSSM Higgs physics at higher orders, for both 
CP-conserving and CP-violating cases, in \cite{Heinemeyer:2004ms}. 
For an extensive discussion of Higgs-sector CP violation, see
the CPNSH report~\cite{Kraml:2006ga}. 

\begin{figure}
\begin{center}
\includegraphics[width=0.48\textwidth,angle=0]{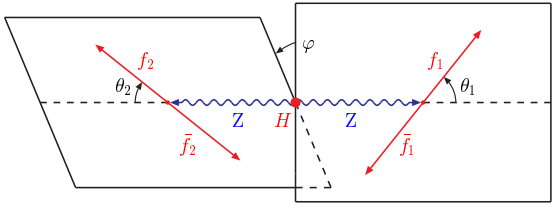}
\caption{Definition of the polar angles ${\theta_i}$ ($i=1,2$) and
the azimuthal angle $\varphi$ for the sequential decay \mbox{$H
\rightarrow Z^{(*)} Z \rightarrow (f_1\bar{f}_1) (f_2\bar{f}_2)$} in
the rest frame of the Higgs boson.}
\label{gmmm_angles}
\end{center}
\end{figure}

\begin{figure}
\begin{center}
\includegraphics[width=0.46\textwidth,angle=0]{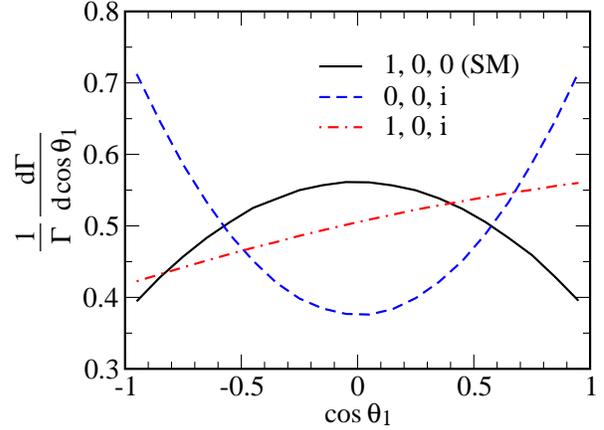}
\caption{The normalized differential width for \mbox{$H \rightarrow
Z Z \rightarrow (f_1\bar{f}_1) (f_2\bar{f}_2)$} with respect to
the cosine of the fermion's polar angle. The solid (black) curve shows
the SM ($a=1$, $b=c=0$) while the dashed (blue) curve is a
pure CP-odd state ($a=b=0$, $c=i$). The dot-dashed (red) curve is for
a state with a CP violating coupling ($a=1$, $b=0$, $c=i$). One can clearly
see an asymmetry in $\cos \theta_1$ for the CP-violating case.}
\label{gmmm_theta}
\end{center}
\end{figure}

A question that naturally arises is whether and how the CP properties of  the 
Higgs boson(s) can be determined at the LHC. (At the ILC, which is a high-precision 
machine in particular for Higgs physics, this can be done quite well, see \cite{AguilarSaavedra:2001rg} 
and references therein). 
A very promising channel is $H\to ZZ\to 4$\,leptons; cf.\ the contributions by Godbole {\it et al.}, 
Buszello and Marquard, and Bluj in  \cite{Kraml:2006ga}. Were here follow 
Godbole~{\it et al.}~\cite{Kraml:2006ga,Godbole:2007cn}:
The $HZZ$ coupling can be written as in the general form
\begin{eqnarray}
   g_{HZZ} &\sim& [\,a\, g_{\mu\nu} +  b \,({k_2}_\mu {k_1}_\nu - k_1 \cdot k_2 g_{\mu \nu}) \nonumber \\ 
   & & \hspace{12mm} +\, c \, \epsilon_{\mu\nu\alpha\beta}\, {k_1}^\alpha {k_2}^\beta\,],
\label{eg:gHZZ}
\end{eqnarray}
up to a factor $ig/(m_Z \cos\theta_W)$,
where $k_1$ and $k_2$ the four-momenta of the two $Z$ bosons. 
The terms associated
with $a$ and $b$ are CP-even, while that associated with $c$ is
CP-odd.  $\epsilon_{\mu \nu\alpha\beta}$ is totally antisymmetric with
$\epsilon_{0123}=1$.  CP violation is be realized if at least one of
the CP-even terms is present (i.e.\ either $a \neq 0$ and/or $b \neq
0$) and $c$ is non-zero. This can be tested through polar and azimuthal 
angular distributions in $H\to Z^{(*)} Z \rightarrow (f_1\bar{f}_1) (f_2\bar{f}_2)$,
c.f.\ Fig.~\ref{gmmm_angles}. 
Denoting the polar angles of the fermions $f_1,f_2$ in the rest frames of the $Z$ bosons by
$\theta_1$ and $\theta_2$, we have e.g.\
\begin{equation}
   \cos \theta_1  = \frac{(\vec{p}_{\bar f_1} - \vec{p}_{f_1}) \cdot 
   (\vec{p}_{\bar f_2} + \vec{p}_{f_2})}{|\vec{p}_{\bar f_1} - \vec{p}_{f_1}| 
   |\vec{p}_{\bar f_2} + \vec{p}_{f_2}|}
\end{equation}
where $\vec{p}_f$ are the three-vectors of the corresponding fermions
with $\vec{p}_{f_1}$ and $\vec{p}_{\bar f_1}$ in their parent $Z$'s
rest frame but $\vec{p}_{f_2}$ and $\vec{p}_{\bar f_2}$ in the Higgs
rest frame, see Fig.~\ref{gmmm_angles}. The angular distribution in
$\theta_i$ ($i=1,2$) for a CP-odd state is $\sim (1+\cos^2\theta_i)$,
corresponding to transversely polarized $Z$ bosons, which is very
distinct from the purely CP-even distribution proportional to
$\sin^2\theta_i$ for longitudinally polarized $Z$ bosons in the large
Higgs mass limit.
$\Im m(c) \neq 0$ will introduce a
term linear in $\cos\theta_i$ leading to a forward-backward
asymmetry. The distribution for $\cos \theta_1$ is shown in
Fig.~\ref{gmmm_theta} for a Higgs mass of $200\,$GeV and a purely
scalar, purely pseudoscalar and CP-mixed scenario. The asymmetry is
absent if CP is conserved (for both CP-odd and CP-even states) but is
non-zero if $\Im m(c) \neq 0$ while simultaneously $a \neq 0$. 
Another probe of CP violation is the azimuthal angular distribution $d\Gamma/d\varphi$ 
with $\varphi$ the angle between the planes of the fermion pairs,  
see Fig.~\ref{gmmm_angles}. For a detailed discussion of 
various distributions and asymmetries sensitive to CP violation 
in $H\to ZZ\to 4$\,leptons, see \cite{Godbole:2007cn}.

\begin{figure}
\includegraphics[width=0.48\textwidth,angle=0]{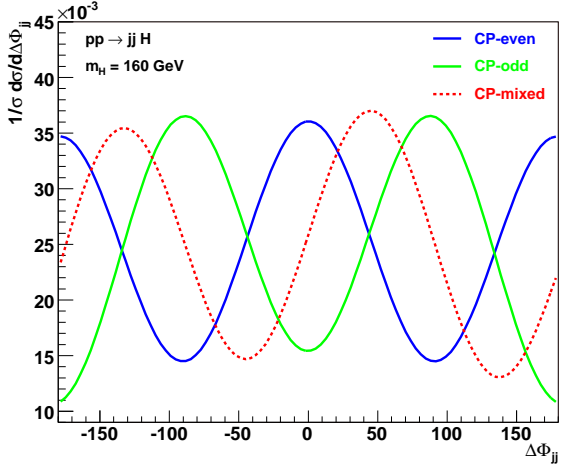}
\caption{Normalized distributions of the jet-jet azimuthal angle difference, 
  for the SM CP-even case ($a_{3}=0$), a pure CP-odd
  ($a_{2}=0$) and a CP-mixed case ($a_{2}=a_{3}\ne 0$); from \cite{Klamke:2007cu}.}
\label{fig:phijjall}       
\end{figure}

Another possibility to test Higgs CP mixing at the LHC are correlations arising 
in the production process.  Here the azimuthal angle correlations between the two
additional jets in $Hjj$ events have emerged as a promising tool~\cite{Plehn:2001nj}.
Higgs boson production in association with two tagging jets, analysed in detail 
in \cite{DelDuca:2006hk}, is mediated by electroweak vector boson fusion and by gluon fusion.  
The latter proceeds through top-quark loops, which induce an effective $Hgg$ vertex.
Writing the $Htt$ Yukawa coupling as 
${\cal L}_Y = y_t H\bar t t +i \tilde y_t A\bar t\gamma_5 t$, 
where $H$ and $A$ denote scalar and  pseudoscalar Higgs fields, 
the tensor structure of the  effective $Hgg$ vertex has the form \cite{Hankele:2006ma,Klamke:2007cu}
\begin{equation}
T^{\mu\nu} = a_2\,(q_1\cdot q_2\, g^{\mu\nu} - 
q_1^\nu q_2^\mu) + a_3\, \varepsilon^{\mu\nu\rho\sigma}q_{1\rho}q_{2\sigma}\,,
\label{eq:Tmunu}
\end{equation}
with 
\begin{equation}
a_2 = \frac{y_t}{y_t^{SM}}\cdot\frac{\alpha_s}{3\pi v} \quad {\rm and}\quad  
a_3 = -\frac{\tilde y_t}{y_t^{SM}}\cdot\frac{\alpha_s}{2\pi v}\,.
\end{equation}
The azimuthal angle correlation of the two jets is hence sensitive to the CP-nature of the 
$Htt$ Yukawa coupling. To resolve interference effects between the CP-even coupling $a_2$ 
and the CP-odd coupling $a_3$ it is, however, important to measure the sign of $\Delta\Phi_{jj}$.
This can be done by defining $\Delta\Phi_{jj}$ as the 
azimuthal angle of the ``away'' jet minus the azimuthal angle of the ``toward'' jet 
with respect to the beam direction~\cite{Hankele:2006ma}.  
The corresponding distributions, for two jets with 
$p_{Tj} > 30$~GeV, $|\eta_j| < 4.5$, and $|\eta_{j_1}-\eta_{j_2}| > 3.0$, 
are shown in Fig.~\ref{fig:phijjall} for three scenarios of CP-even 
and CP-odd Higgs couplings~\cite{Klamke:2007cu}. 
All three cases are well distinguishable, with the maxima in the distributions directly 
connected to the size of the scalar and pseudoscalar contributions, $a_2$ and $a_3$. 

\subsection{Gauginos and sfermions}\label{sec:gauginos}

The CP-violating phases in (\ref{eq:phases}) directly enter the neutralino, chargino, and sfermion 
mass matrices, hence having an important impact on the masses and couplings of these particles. 
This is particularly interesting for the precision measurements envisaged at the ILC. 
The effects of CP phases in measurements of neutralinos, charginos, and sfermions at the ILC have 
been studied in great detail by various groups; see below as well as references in \cite{Ibrahim:2007fb,MoortgatPick:2005cw}. They fall into two different classes. 
On the one hand, there are CP-even observables: spartice masses, cross sections, 
branching ratios, etc.. If measured precisely enough, they allow for a parameter 
determination, either analytically \cite{Choi:2000ta,Kneur:1998gy} or through a global fit \cite{Bartl:2003pd}. 
Beam polarization \cite{MoortgatPick:2005cw} is essential, 
but some ambiguities in the phases always remain. 
We do not discuss this in more detail here.
On the other hand, there are CP-odd (or T-odd) observables, e.g.\ rate asymmetries or  
triple-product asymmetries, which are a direct signal of CP violation. 
Indeed the measurement of CP-odd effects is necessary to prove that CP is violated, and to 
determine the model parameters, including phases, in an unambiguous way.

\begin{figure}
\begin{center}
\includegraphics[width=0.48\textwidth,angle=0]{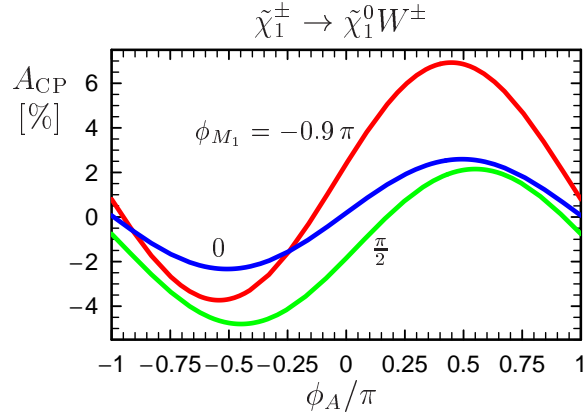}
\end{center}
\caption{The dependence of $A_{CP}^{\tilde\chi^\pm}$ on $\phi_A$, and 
various values of $\phi_{M_1}$, from \cite{Eberl:2005ay}.}
\label{fig:eberl}
\end{figure}

An expample for a rate asymmetry is the chargino decay into a neutralino and 
a $W$ boson, $\tilde\chi_i^\pm \to \tilde\chi_j^0 W^\pm$. 
Here, non-zero phases can 
induce an asymmetry between the decay rates of $\tilde\chi_i^+$ and $\tilde\chi_i^-$, 
\begin{equation}
A_{\rm CP} = 
  \frac{ \Gamma(\tilde\chi_i^+ \to \tilde\chi_j^0 W^+) - \Gamma(\tilde\chi_i^- \to \tilde\chi_j^0 W^-)}{
           \Gamma(\tilde\chi_i^+ \to \tilde\chi_j^0 W^+) + \Gamma(\tilde\chi_i^- \to \tilde\chi_j^0 W^-)}\,, 
\end{equation}
through absorptive parts in the one-loop corrections~\cite{Eberl:2005ay}. 
Figure~\ref{fig:eberl} shows the dependence of $A_{\rm CP}$ on $\phi_{A}\equiv \phi_{t,b,\tau}$ for  
$M_2 = 500$~GeV, $|\mu| = 600$~GeV, 
$|A| = 400$~GeV, $M_{\tilde Q} = 400$~GeV,   and various $\phi_{M_1}$.
$A_{\rm CP}$ has its maximum at $|\phi_{A}| \sim \pi/2$ and is larger at 
large negative values of the phase of $M_1$. 
The obvious advantage of such a rate asymmetry is that it can be measured 
in a `simple' counting experiment.
Analogous asymmetries have been computed for $H^\pm$ 
in \cite{Christova:2002ke,Christova:2006fb,Frank:2007ca}.
Ref.~\cite{Christova:2006fb} also discusses CP-violating forward-backward asymmetries.

\begin{figure*}
\begin{center}
\includegraphics[width=0.9\textwidth,angle=0]{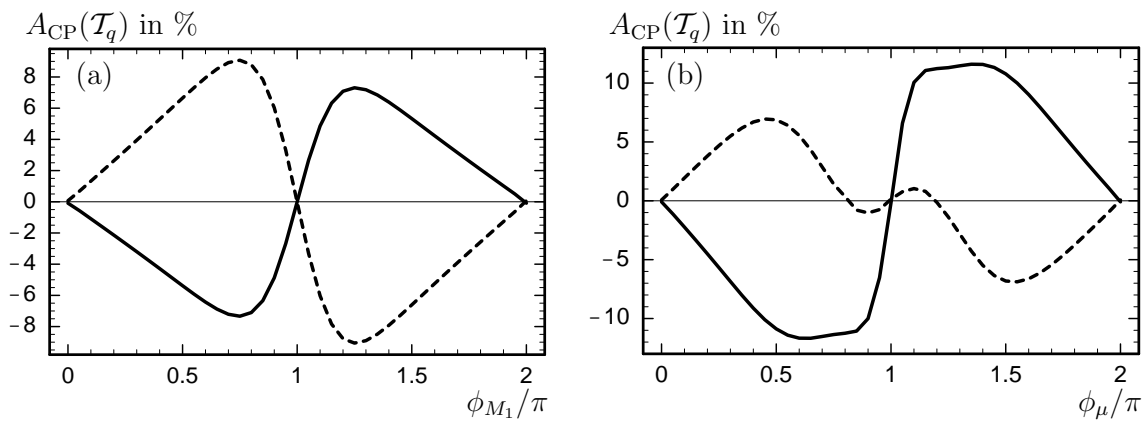}
\end{center}
\caption{CP asymmetry $A_{\rm CP}(\mathcal{T}_q)$ for $e^+e^- \to \tilde{\chi}^+_1\tilde{\chi}^-_2$
with subsequent decay $\tilde{\chi}^+_1 \to \tilde{\chi}^0_1 \bar{s} c$ 
for $\sqrt{s}=500$~GeV and beam polarizations $(P_{e^-},P_{e^+}) = (-0.8, +0.6)$ (solid),
$(P_{e^-},P_{e^+}) = (+0.8, -0.6)$ (dashed); the parameters are 
$M_2=280$ GeV, $|\mu|=200$ GeV, $\tan\beta=5$, with 
$\phi_\mu=0$ in (a) and $\phi_1=0$ in (b); from \cite{Bartl:2006yv}.}
\label{fig:bartl}
\end{figure*}

Triple-product asymmetries rely on spin correlations between sparticle production and decay 
processes. They have been computed for neutralino \cite{Kizukuri:1990iy,Choi:1999cc,Barger:2001nu,Bartl:2003tr,Choi:2004rf,AguilarSaavedra:2004dz,Bartl:2004jj,Choi:2005gt} and chargino 
\cite{Bartl:2004vi,Kittel:2004kd,Bartl:2006yv} production in $e^+e^-$ 
followed by two- or three-body decays. 
Let me take the most recent work \cite{Bartl:2006yv} on chargino-pair production with 
subsequent three-body decay as an illustrative example. 
The processes considered are 
$e^+e^-\rightarrow \tilde{\chi}^+_1\tilde{\chi}^-_j$ ($j=1,2$) at a linear collider with 
longitudinal beam polarizations, followed by three-body decays of the $\tilde{\chi}^+_1$, 
\begin{equation}\label{eq:chidecay}
  \tilde{\chi}^+_1 \to \tilde{\chi}^0_1~\nu~\ell^+ \quad {\rm or} \quad
  \tilde{\chi}^+_1 \to \tilde{\chi}^0_1~\bar{s}~c \,,
\end{equation}
where $\ell=e,\mu$. 
It is assumed that the momenta $\vec{p}_{\tilde\chi_1^+}$, 
$\vec{p}_{\ell}$, $\vec{p}_{c}$ and $\vec{p}_{s}$ of the 
associated particles can be measured or reconstructed.
The relevant triple products are: 
\begin{eqnarray} 
  \label{tpdeflep}
  \mathcal{T}_\ell&=&\vec{p}_{\ell^+} \cdot (\vec{p}_{e^-} \times \vec{p}_{\tilde{\chi}^+_1})\,,\\
  \label{tpdefhad}
  \mathcal{T}_q&=&\vec{p}_{\bar{s}}\cdot (\vec{p}_{c}\times\vec{p}_{e^-})\,.
\end{eqnarray}
Note that $\mathcal{T}_\ell$, relates momenta of initial, intermediate and final particles, 
whereas $\mathcal{T}_q$, uses only momenta from the initial and final states. Therefore, 
both triple products depend in a different way on the 
production and decay processes. From $\mathcal{T}_{\ell,q}$ one can 
define T-odd asymmetries 
\begin{equation} \label{eq:Todd}
A_T(\mathcal{T}_{\ell,q})=
\frac{N[\mathcal{T}_{\ell,q}> 0]-N[\mathcal{T}_{\ell,q} < 0]}
{N[\mathcal{T}_{\ell,q}> 0]+N[\mathcal{T}_{\ell,q} < 0]}\,,
\end{equation}
where $N[\mathcal{T}_{\ell,q} > (<)~ 0]$ is the number
of events for which $\mathcal{T}_{\ell,q}> (<) ~0$.
A genuine signal of CP violation is obtained by 
combining $A_T(\mathcal{T}_{\ell,q})$ with the corresponding asymmetry
$\bar{A}_T(\mathcal{T}_{\ell,q})$ for the charge-conjugated processes:
\begin{equation}\label{ACP}
  A_{\rm CP}(\mathcal{T}_{\ell,q})=
  \frac{A_T(\mathcal{T}_{\ell,q})-\bar{A}_T(\mathcal{T}_{\ell,q})}{2}\,.
\end{equation}
Figure~\ref{fig:bartl} shows the phase dependence of $A_{\rm CP}(\mathcal{T}_q)$  
for $e^+e^- \to \tilde{\chi}^+_1\tilde{\chi}^-_2$ followed by $\tilde{\chi}^+_1 \to \tilde{\chi}^0_1 \bar{s} c$ 
for $\sqrt{s}=500$~GeV and polarized $e^{\pm}$ beams. 
The authors conclude that $A_{\rm CP}(\mathcal{T}_q)$ can be probed at the $5 \sigma$ level
in a large region of the MSSM parameter space, while $A_{\rm CP}(\mathcal{T}_\ell)$ has a 
somewhat lower sensitivity.

\subsection{EDM constraints}\label{sec:edms}

Let us next discuss the EDM constraints in some more detail.  
The constraints eq.~(\ref{eq:edmconstr}), especially the one on $d_{\rm Tl}$, 
translate into a tight bound on the electron EDM,  
\begin{equation}
   |d_e|<1.6\times 10^{-27}\, e\, {\rm cm}.
\end{equation}
Setting all soft breaking parameters in the selectron and gaugino sector 
equal to $M_{\rm SUSY}$, one can derive a simplified formula for the one-loop 
contributions \cite{Falk:1999tm} 
\begin{eqnarray}
  d_e &=& f_{\rm S}^{} \Big[ \Big(\frac{5g_2^2}{24}+ \frac{g_1^2}{24}\Big) 
                  \sin [{\rm Arg} (\mu M_2)] \tan\beta \nonumber\\
 & & \qquad +\; \frac{g_1^2}{12}\sin [{\rm Arg} (M_1^* A_e)]\big] \,,
\label{simplified}
\end{eqnarray}
where $f_{\rm S}=(em_e)/(16 \pi^2 M_{\rm SUSY}^2)$, and ${\rm Arg}(B\mu)=0$ by convention.
Note the $\tan\beta$ enhancement of the first term. 
It is the main reason why the phase of $\mu$ is 
more severely constrained than the phases of the $A$ parameters. 
The phases of the third generation, $\phi_{t,b,\tau}$, only enter the EDMs at the 
two-loop level. However, there can be a similar $\tan\beta$ enhancement for these 
two-loop contributions \cite{Pilaftsis:1999bq}, so they have to be taken into account as well. 

\begin{figure}\begin{center}
  \includegraphics[width=0.42\textwidth,angle=0]{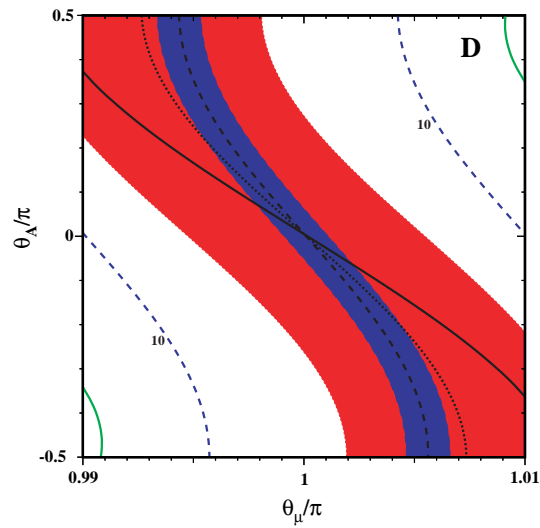}
  \end{center}
  \caption{The Tl (blue dashed) and neutron (red dotted)
 EDMs relative to their respective experimental limits in the 
 $\theta_\mu, \theta_A$ plane for benchmark point D of \cite{Olive:2005ru}.  
 Inside the shaded regions, the EDMs are less than
 or equal to their experimental bounds. Each of the EDMs vanish along the black contour
 within the shaded region; from \cite{Olive:2005ru}.}
\label{fig:edmD}
\end{figure}

Indeed, the EDM constraints pose a serious problem in the general MSSM: 
for $O(100)$~GeV masses and $O(1)$ phases, the EDMs are typically three(!) 
orders of magnitude too large \cite{ellis,buch,pol,dug}. Some efficient suppression 
mechanism is needed to satisfy the experimental bounds. 
The possibilities include 
\begin{itemize}
\item small phases, 
\item heavy sparticles  \cite{nath,ko,ckn,Falk:1995fk},
\item accidental cancellations \cite{fo1,nath2,ca1,ca2,ca3,ca4,ca5},
\item flavour off-diagonal CP phases \cite{Abel:2001vy},
\item lepton flavour violation \cite{Bartl:2003ju}.
\end{itemize}
Detailed analyses of the EDM constraints have recently 
been performed e.g.\ in \cite{Pospelov:2005pr,Olive:2005ru,Ayazi:2007kd}.
As example that large phases can be in agreement with the current EDM limits,
Fig.~\ref{fig:edmD} shows the results for the CMSSM benchmark point D of \cite{Olive:2005ru}, 
which has $(m_{1/2}, m_0, \tan \beta) = (525, 130, 10)$. The strongest constraint 
comes from the EDM of Tl; that of Hg is not shown because it is satisfied over the 
whole plane.  As can be seen, for this benchmark point there is no limit to $\theta_A$, 
while $|\theta_\mu - \pi| \le 0.065 \pi$.

\subsection{Neutralino relic density}\label{sec:relic}

\begin{figure*}
  \includegraphics[width=0.48\textwidth,angle=0]{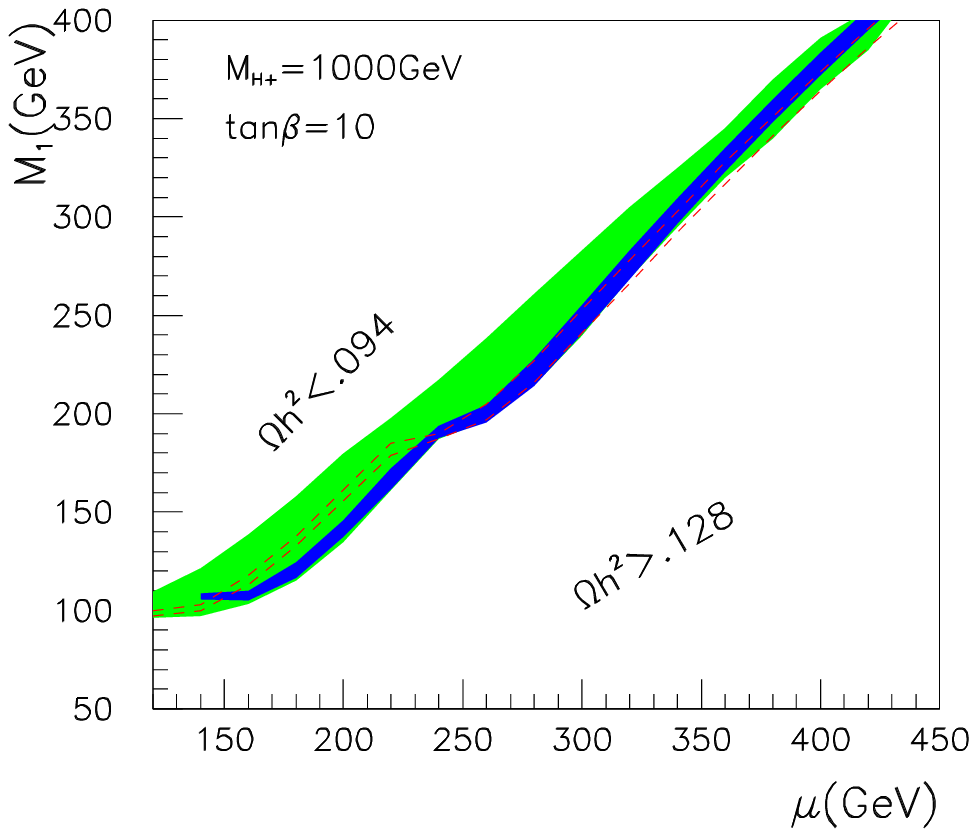}\quad
  \includegraphics[width=0.48\textwidth,angle=0]{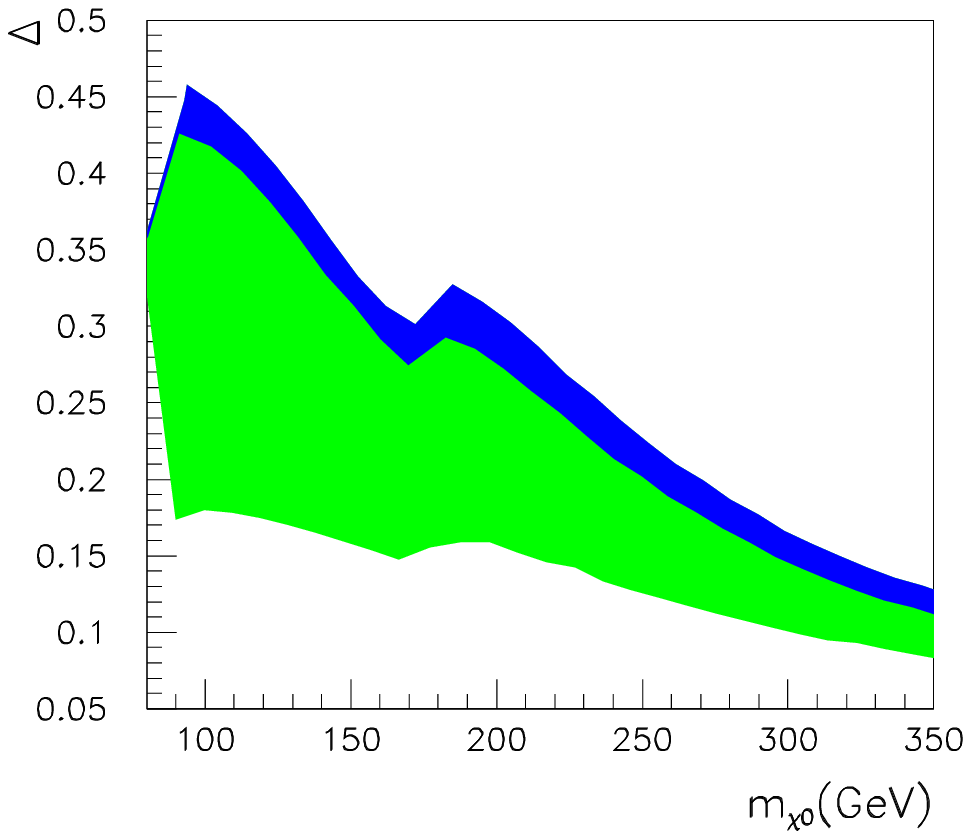}
  \caption{Left: The $2\sigma$ WMAP bands in the $M_1$--$\mu$
           plane for $\tan\beta=10$, $m_{H^+}=M_S=A_t=1$~TeV,
           for all phases zero (blue/dark grey band),
           for $\phi_\mu=180^\circ$ (or $\mu<0$) and all other phases zero
           (dashed red lines)
           and for arbitrary phases (green/light grey band).
           Right: The corresponding relative mass difference
               $\Delta\equiv (m_{\tilde\chi^\pm_1}-m_{\tilde\chi^0_1})/m_{\tilde\chi^0_1}$
               as function of $m_{\tilde\chi^0_1}$ for all phases zero (blue/dark grey band)
               and for arbitrary phases (green/light grey band). From \cite{Belanger:2006qa}.}
\label{fig:m1mumh1000}
\end{figure*}

If the $\tilde\chi^0_1$ is the lightest supersymmetric particle (LSP) and stable, it is a very 
good cold dark matter candidate.
In the framework of thermal freeze-out, its relic density is $\Omega h^2\sim 1/\langle\sigma_Av\rangle$, 
where $\langle\sigma_Av\rangle$ is the thermally averaged annihilation cross section summed over
all contributing channels. These channels are: 
annihilation of a bino LSP into fermion pairs through $t$-channel
sfermion exchange in case of very light sparticles;
annihilation of a mixed bino-Higgsino or bino-wino LSP into gauge
boson pairs through $t$-channel chargino and neutralino exchange,
and into top-quark pairs through $s$-channel $Z$ exchange;
and annihilation near a Higgs resonance (the so-called Higgs funnel); 
and finally coannihilation processes with sparticles that are close
in mass with the LSP. Since the neutralino couplings to other (s)particles 
sensitively depend on CP phases, the same can be expected for 
$\langle\sigma_Av\rangle$ and hence $\Omega h^2$.

The effect of CP phases on the neutralino relic density was considered in \cite{Falk:1995fk,Gondolo:1999gu,Nihei:2004bc,Argyrou:2004cs,Balazs:2004ae,Gomez:2005nr,Choi:2006hi,Cirigliano:2006dg,Lee:2007ai}, although only for specific cases. The first general analysis, 
(i) including all annihilation and coannihilation processes and (ii) separating the phase dependece of 
the couplings from pure kinematic effects, was done in \cite{Belanger:2006qa}. 

It was found that modifications in the couplings due to non-trivial CP phases 
can lead to variations in the neutralino relic density of up to an order of magnitude. 
This is true not only for the Higgs funnel but also for other scenarios, 
like for instance the case of a mixed bino-higgsino LSP. 
Even in scenarios which feature a modest phase
dependence once the kinematic effects are singled out, the
variations in $\Omega h^2$ are comparable to (and often much
larger than) the $\sim 10\%$ range in $\Omega h^2$ of the WMAP
bound. 
Therefore, when aiming at a precise prediction of the neutralino relic
density from collider measurements, it is clear that one does not only
need precise sparticle spectroscopy but one also has to precisely measure
the relevant couplings, including possible CP phases.

This is illustrated in Fig.~\ref{fig:m1mumh1000}, which shows 
the regions where the relic density is in agreement with the $2\sigma$ WMAP bound,
$0.0945 < \Omega_{\rm CDM} h^2 < 0.1287$,
for the case of a mixed bino-higgsino LSP.
When all phases are zero, only the narrow
blue (dark grey) band is allowed. When allowing all phases to vary arbitrarily, while still 
satisfying the EDM constraints, the allowed band increases to the 
the green (light grey) region. In the $|M_1|$--$|\mu|$ plane (left panel), 
the allowed range for $\mu$ increases roughly from $\delta\mu\sim 10$~GeV
to $\delta\mu\sim 50$~GeV for a given $|M_1|$. In terms of relative mass differences (right panel) 
this means that in the CP-violating case much smaller $\tilde\chi^\pm_1$--$\tilde\chi^0_1$
mass differences can be in agreement with the WMAP bound
than in the CP-conserving case.

\section{Conclusions} 

The observed baryon asymmetry of the Universe necessiates new sources of CP 
violation beyond those of the SM. I this talk, I have discussed effects of such new 
CP phases, focussing on the case of the MSSM. The topics covered  include 
CP-mixing in the Higgs sector and its measurement at the LHC, 
CP-odd observables in the gaugino sector at the ILC, 
EDM constraints, and the neutralino relic density.     
Each topic was discussed by means of some recent example(s) from the literature.
For a more extensive discussion, in particular of topice that could not be covered here, 
I refer the reader to the recent review by Ibrahim and Nath~\cite{Ibrahim:2007fb}.


\end{document}